\documentclass[11pt,twoside]{article}
\usepackage{tikz}
\usepackage{asp2010}
\usepackage[squaren]{SIunits}
\usepackage{graphicx}
\usepackage{epstopdf}

\usetikzlibrary{intersections,positioning,backgrounds,fit,matrix,shapes,
                arrows,calc,decorations.pathmorphing,decorations.text,
                shadows}

\resetcounters

\include{O4-2_tikz}

\begin{document}

\title{Managing Hardware Configurations and Data Products for the Canadian Hydrogen Intensity Mapping Experiment}

\author{
Adam D. Hincks$^{1}$ and J. Richard Shaw$^{2}$ for the CHIME Collaboration
\affil{$^1$Department of Physics and Astronomy, University of British Columbia, 6224 Agricultural Rd. Vancouver BC V6T 1Z1, Canada}
\affil{$^2$The Canadian Institute for Theoretical Astrophysics, 60 St George St, Toronto ON M5S 3H8, Canada}
} % End author

\begin{abstract}
The Canadian Hydrogen Intensity Mapping Experiment (CHIME) is an ambitious new radio telescope project for measuring cosmic expansion and investigating dark energy. Keeping good records of both physical configuration of its 1280 antennas and their analogue signal chains as well as the $\sim$100\,TB of data produced daily from its correlator will be essential to the success of CHIME. In these proceedings we describe the database-driven software we have developed to manage this complexity.
\end{abstract}

\section{Introduction}

The Canadian Hydrogen Intensity Mapping Experiment (CHIME) will produce a 3D map of large scale cosmic structure at redshifts $0.8 < z < 2.5$ by measuring the 21\,cm emission of neutral hydrogen. This will lead to a precise determination of the expansion history of the Universe and shed light on the nature of dark energy \citep{bandura/etal:2014, newburgh/etal:2014}.

CHIME will be a complex experiment. It consists of five parallel, $20\,\metre\times100\,\metre$ cylindrical reflectors, each populated with 256 dual-polarisation antennas along its focal line. The signal chain of each polarisation includes two amplifiers and more than $60\,\meter$ of coaxial cable prior to digitisation. Together with the housekeeping for monitoring amplifier temperatures, CHIME uses close to 10,000 components and over 160 km of cable. All antenna signals are digitised at $800\,\mega\hertz$, Fourier transformed to 1024 frequency channels and then correlated. The total raw output after co-adding the correlated products in $10\,\second$ intervals is about $100\,\mathrm{TB}$ per day.
% 256 * 5 antennas
% 256 * 5 * 2 LNA's, FLA's
% 256 * 5 * 2 * 2 SMA coax
% 256 * 5 * 2 thermometers
% Size in TB: (256*5*2)^2 * 2 * 1024 * 360 * 24 / 1024^4

Consideration of the above reveals two organisational challenges. First, the configuration of the analogue components deployed on CHIME needs to be recorded, and tracked when it changes. Second, the large flow of data needs to be managed efficiently and data products need to be readily associated with particular physical configurations. In this paper, we describe a database-driven approach to meeting these challenges.

\section{Tracking Physical Configurations}
\label{sec:layout}

\begin{figure}[tb]
  \includegraphics[width=\textwidth]{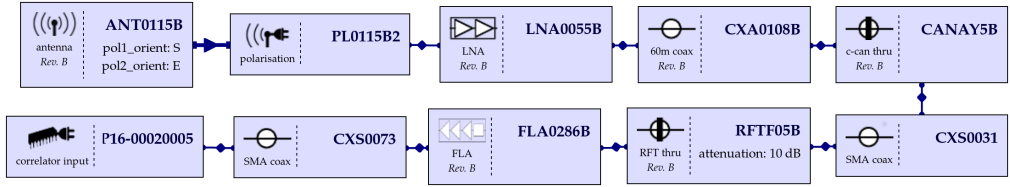}
  \caption{A graph of the path from one antenna output to the correlator produced by the web interface. The thick line with an arrow indicates a permanent connection.}
\label{fig:layout}
\end{figure}

Mathematically, the physical configuration of components deployed on CHIME is a graph. Its nodes are components, which can have changeable properties, including the state of a component's deployment (such as the position of an antenna on a focal line), its calibration data and its production/usage history. Each component is labelled with a serial number, the majority of which are barcoded to enable machine identification. The edges of the graph are connections between components. Finally, the graph itself can have global properties, such as a tag labelling a stable configuration or an event like a period of known radio contamination. Fig.~\ref{fig:layout} shows a section of a CHIME graph.

As the experiment proceeds, the configuration is far from static. Thus, every time a component is added, removed, connected or disconnected from another component, or has its properties altered, the graph changes.

\begin{figure}[tb]
  \centering\includegraphics{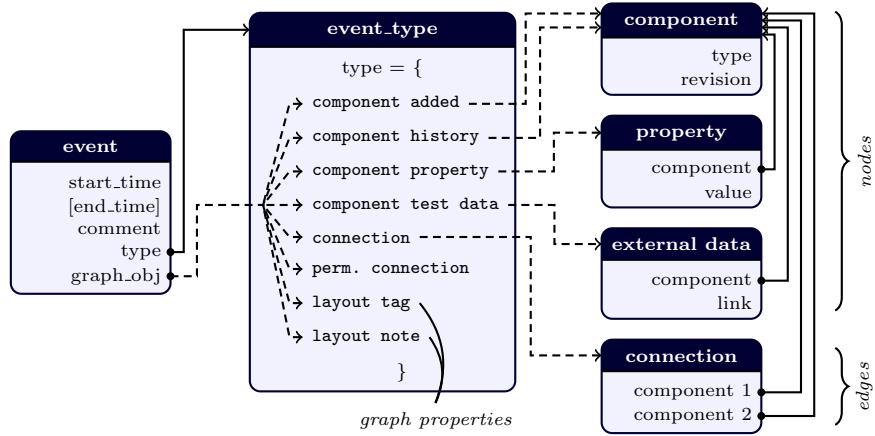}
  \caption{A simplified schematic of the database architecture for tracking  CHIME configurations. {\texttt Events} are polymorphically related to specific graph objects as per the \texttt{event type}. Graph objects are nodes, edges and node/graph properties.}
  \label{fig:polymorphic}
\end{figure}

The CHIME graph is tracked in a \texttt{MySQL} database by recording \textit{events}. An event has a timestamp for its beginning and, optionally, a timestamp for its end. A set of \textit{event types} define the polymorphic relationships between the event table and \textit{graph objects}, where graph objects are nodes, connections and node or graph properties. Fig.~\ref{fig:polymorphic} is a simplified illustration of this design. The event table can be queried to select all graph objects active at a given time to determine CHIME's configuration graph at that time.

We have written a \texttt{python} module for managing the database. We use the \texttt{peewee} object-relational mapping (ORM) package to query the database,\footnote{http://peewee.readthedocs.org} adding a layer of functionality that ensures that the database integrity for our event-driven model is maintained. For example, the \texttt{python} module forbids making a connection if it would overlap with (and therefore duplicate) an existing connection event at that time.

The \texttt{python} module can also query the database to construct our configuration graph at a specified time. The graph is provided to the user as a \texttt{NetworkX} object,\footnote{http://networkx.github.io} allowing the extensive tools \texttt{networkx} provides for analysing the graph, such as finding nearest neighbours. On top of \texttt{networkx} we have added a layer of functionality to provide ``convenience'' methods for common operations on the graph.

For entry of multiple events, we have created a simple, text-based mark-up language in which the making and severing of connections and the assignment of node properties can be represented. The markup language is designed such that a series of connections scanned with a barcode reader into a text file can be read to the \texttt{python} module with minimal (or sometimes no) further editing.

We are also developing a web interface for querying the database. It is coded in \texttt{PHP} and makes use of \texttt{Propel} for the ORM interface to the database.\footnote{http://propelorm.org/} (Some complex queries are executed by \texttt{CGI} scripts utilising our \texttt{python} module.) The web interface provides user-friendly access to the database, including useful visualisation tools. For instance, it can draw graphical representations of portions of our graphs using \texttt{jsPlumb},\footnote{https://jsplumbtoolkit.com/} an example of which is shown in Fig.~\ref{fig:layout}.

\section{Management of Data Products}
\label{sec:alpenhorn}

Output from the CHIME correlator is organised into folders called \textit{acquisitions}. Each acquisition, which represents an uninterrupted period of data-taking, consists of one or more datafiles of roughly three hours' duration, as well as a log file. 

In the same database described in Section~\ref{sec:layout}, we record the name, size and MD5 hash of each file in an \texttt{ArchiveFile} table. Archive files are associated an \texttt{ArchiveAcq} table denoting their acquisition. Acquisitions are associated with a \texttt{CorrAcqInfo} table and correlator files that contain raw data (as opposed, say, to log files) are associated with \texttt{CorrFileInfo} table. These provide metadata that are useful for querying, and are placed in separate tables to allow other types of acquisition, such as housekeeping, to have their own metadata tables. Crucially, the \texttt{CorrFileInfo} table records the beginning and duration of an acquisition file. Hence, with a simple query, a user can discover which configuration graph (see Section~\ref{sec:layout}) was in place during acquisition. A \texttt{python} class has been developed to return all the data to the user based on common search parameters.

The same files can exist on multiple \textit{storage nodes}, both because we want backup copies and because data are analysed off-site at the SciNet cluster in Toronto. An \texttt{ArchiveFileCopy} table associated with the \texttt{ArchiveFile} table tracks all copies of a single file.

\begin{figure}[tb]
  \centering\includegraphics{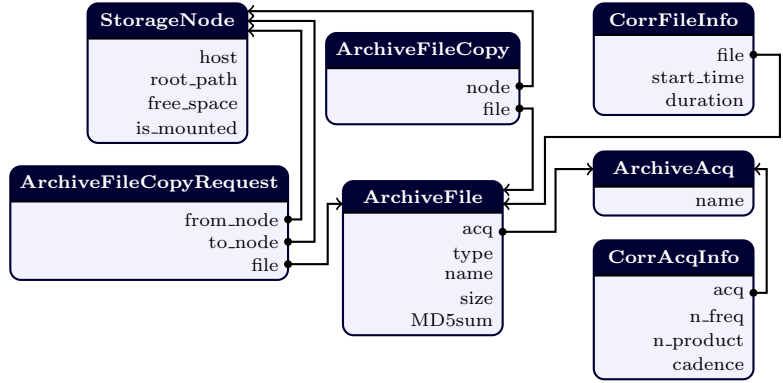}
  \caption{A simplified schematic of the relational database architecture for managing data products. See text for details.}
  \label{fig:di}
\end{figure}

A \texttt{python} script called \texttt{alpenhorn} manages the database tables described above. Every computer host that has a storage node runs an instance of this script as a daemon. On the acquisition computer, \texttt{alpenhorn} watches for new files and registers them in the database. Copies are requested by adding a row to a \texttt{ArchiveFileCopyRequest} table; this table is normally updated by a \texttt{cron} job that checks for files do not yet exist off-site. Instances of \texttt{alpenhorn} running on off-site hosts respond to an \texttt{ArchiveFileCopyRequest} by \texttt{rsyncing} the file from the appropriate storage node, checking the MD5 hash, and then updating the \texttt{ArchiveFileCopy} table. To clear a file copy from a storage node, its entry in \texttt{ArchiveFileCopy} is marked for removal. It will be deleted by the \texttt{alpenhorn} instance running on that storage node---but only if two copies exist elsewhere.

\section{Current Status}

Construction on the full CHIME instrument has begun. In the meantime, we have been working with a ``pathfinder'' instrument of about one-tenth the scale. Most aspects of the database software described in this paper are already being used on the pathfinder. This has been invaluable for introducing improvements---for example, the event-driven model for recordings graphs has replaced an unwieldy, static model.

The management of data products works well, but was put in place before the event-driven graph model was developed. A future implementation might benefit from having the existence of data products fully integrated into the event model.

\acknowledgements We thank Michael Nolta for sharing code that informed the design described in Sec.~\ref{sec:alpenhorn}. We are grateful for to the staff of the Dominion Radio Astrophysical Observatory, operated by the National Research Council Canada. We acknowledge support from the Canada Foundation for Innovation and the Natural Sciences and Engineering Research Council of Canada.

\bibliographystyle{asp2010}
\bibliography{O4-2}

\end{document}